\documentclass[reqno,11pt]{amsart}
\usepackage{amsmath, latexsym, amsfonts, amssymb, amsthm, amscd}
\usepackage{graphics,epsf,psfrag}
\numberwithin{equation}{section}

\newtheorem{theorem}{Theorem}[section]

\newtheorem{rem}[theorem]{Remark}

\newcommand{\ind}{\mathbf{1}}

\newcommand{\R}{\mathbb{R}}
\newcommand{\Z}{\mathbb{Z}}
\newcommand{\N}{\mathbb{N}}
\renewcommand{\tilde}{\widetilde}

\newcommand{\cL}{{\ensuremath{\mathcal L}} }

\newcommand{\cD}{{\ensuremath{\mathcal D}} }

\newcommand{\bP}{{\ensuremath{\mathbf P}} }
\newcommand{\bE}{{\ensuremath{\mathbf E}} }

\DeclareMathSymbol{\leqslant}{\mathalpha}{AMSa}{"36} 
\DeclareMathSymbol{\geqslant}{\mathalpha}{AMSa}{"3E} 
\DeclareMathSymbol{\eset}{\mathalpha}{AMSb}{"3F}     
\newcommand{\dd}{\,\text{\rm d}}             
\newcommand{\suptwo}[2]{\sup_{\substack{#1 \\ #2}}} 
\newcommand{\sumtwo}[2]{\sum_{\substack{#1 \\ #2}}} 

\newcommand{\bbE}{{\ensuremath{\mathbb E}} }

\newcommand{\bbP}{{\ensuremath{\mathbb P}} }

\newcommand{\go}{\omega}

\newcommand{\cf}{\mathcal N_N}
\newcommand{\bPo}{\bP_{N,\go}^{\beta,h}}
\newcommand{\Zno}{Z_{N,\go}^{\beta,h}}
\newcommand{\Hno}{\mathcal H_{N,\go}^{\beta,h}}

\makeatletter
\def\captionfont@{\footnotesize}
\def\captionheadfont@{\scshape}

\long\def\@makecaption#1#2{%
  \vspace{2mm}
  \setbox\@tempboxa\vbox{\color@setgroup
    \advance\hsize-6pc\noindent
    \captionfont@\captionheadfont@#1\@xp\@ifnotempty\@xp
        {\@cdr#2\@nil}{.\captionfont@\upshape\enspace#2}%
    \unskip\kern-6pc\par
    \global\setbox\@ne\lastbox\color@endgroup}%
  \ifhbox\@ne 
    \setbox\@ne\hbox{\unhbox\@ne\unskip\unskip\unpenalty\unkern}%
  \fi
  \ifdim\wd\@tempboxa=\z@ 
    \setbox\@ne\hbox to\columnwidth{\hss\kern-6pc\box\@ne\hss}%
  \else 
    \setbox\@ne\vbox{\unvbox\@tempboxa\parskip\z@skip
        \noindent\unhbox\@ne\advance\hsize-6pc\par}%
\fi
  \ifnum\@tempcnta<64 
    \addvspace\abovecaptionskip
    \moveright 3pc\box\@ne
  \else 
    \moveright 3pc\box\@ne
    \nobreak
    \vskip\belowcaptionskip
  \fi
\relax
}
\makeatother
\def\writefig#1 #2 #3 {\rlap{\kern #1 truecm
\raise #2 truecm \hbox{#3}}}

\newcommand{\tf}{\textsc{f}}

\begin{document}

\title[Critical properties of depinning transitions]
{Critical properties and finite--size estimates for the depinning
transition of
  directed  random polymers}

\author{Fabio Lucio Toninelli}
\address{
Laboratoire de Physique, UMR-CNRS 5672, ENS Lyon, 46 All\'ee d'Italie, 
69364 Lyon Cedex 07, France
\hfill\break
\phantom{br.}{\it Home page:}
{\tt http://perso.ens-lyon.fr/fabio-lucio.toninelli}}
\email{fltonine@ens-lyon.fr}
\date{\today}

\begin{abstract}
  We consider models of directed random polymers interacting with a
  defect line, which are known to undergo a pinning/depinning (or
  localization/delocalization) phase transition. We are interested in
  critical properties and we prove, in particular, finite--size upper
  bounds on the order parameter (the {\em contact fraction}) in a
  window around the critical point, shrinking with the system size.
  Moreover, we derive a new inequality relating the free energy $\tf$
  and an annealed exponent $\mu$ which describes extreme fluctuations
  of the polymer in the localized region.  For the particular case of
  a $(1+1)$--dimensional interface wetting model, we show that this
  implies an inequality between the critical exponents which govern
  the divergence of the disorder--averaged correlation length and of
  the typical one.  Our results are based on on the recently proven
  smoothness property of the depinning transition in presence of
  quenched disorder and on concentration of measure ideas.  \\ \\ 2000
  \textit{Mathematics Subject Classification: 82B27, 82B44, 82B41 } \\
  \\ \textit{Keywords: Directed Polymers, Pinning and Wetting Models,
    Copolymers, Depinning Transition, Finite--Size Estimates,
Concentration of Measure, Typical and Average Correlation Lengths.}
\end{abstract}

\maketitle

\section{Introduction}

Directed polymers interacting with a one--dimensional defect line are
quite rich in physical and biological applications, and lately have
started to attract much attention also in the mathematical literature
\cite{cf:AS,cf:petrelis}, \cite{cf:G}--\cite{cf:GTloc}. In
particular, they are an ideal framework to model $(1+1)$--dimensional
interface wetting phenomena \cite{cf:DHV}, the problem of depinning of
flux lines from columnar defects in type--II superconductors
\cite{cf:Nelson} and the denaturation transition of DNA in the
Poland--Scheraga approximation \cite{cf:KMP}.  In many situations, the
polymer--defect interaction is neither homogeneous nor periodic along
the line.  This corresponds for instance to the presence of impurities
on the wall in the case of the wetting problem, and to the
non--periodic arrangement of base pairs A--T, G--C along DNA
sequences. Therefore, one resorts very naturally to quenched
disordered models.

The interplay between the ({\em energetic}) pinning effect, which
tends to keep the polymer close to the defect line, and the ({\em
  entropic}) depinning one, favoring configurations which wander away
from the line, is responsible for a non--trivial pinning/depinning (or
localization/delocalization) phase transition. The depinned and pinned
phases are characterized by a different behavior of the order
parameter, the {\em contact fraction}, which is essentially the
density of polymer--defect contacts along the line. In the pinned
phase, the contact fraction stays positive in the thermodynamic limit,
while it vanishes in the interior of the depinned phase (finite--size
estimates of the latter statement can be found in \cite{cf:GT}).  A
very interesting problem is to understand what happens at the {\em
  critical line} separating the two phases. Recently, with G. Giacomin
we proved that, as soon as disorder is present, the contact fraction
vanishes continuously when the critical line is approached from the
pinned region \cite{cf:GT05}. This is in striking contrast with the
situation in pure (i.e., non--disordered) pinning models, where the
transition can be either of first or of higher order, depending for
instance on the space dimension.  Given this result, it is very
natural to investigate how fast the contact fraction vanishes with
system size, {\em at} the critical line or in a small critical window
around it.  This question is addressed in Theorem \ref{th:fss} of the
present paper, where it is shown for instance that, in the disordered
situation, the contact fraction is at most of order $N^{-1/3}\log N$
at criticality.

Inside the localized region, the length of the maximal excursion of
the polymer (i.e., of the longest portion of the polymer without
contacts with the defect line) is $(\log N)/\mu$ \cite{cf:AZ,cf:GTloc},
where $\mu$ is a certain annealed exponent (cf. Section
\ref{sec:versus} for its definition) and $N$ is the total length
of the polymer. When the critical line is approached $\mu$ tends to
zero, as well as the free energy $\tf$.  In Theorem \ref{th:bounds}
we prove an inequality which essentially relates the critical
exponents which govern the vanishing of $\mu$ and $\tf$ at the
critical line. This inequality is interesting also because, in the
particular case of a $(1+1)$--dimensional wetting model, we prove in
Theorem \ref{th:correlazioni} that $\tf^{-1}$ and $\mu^{-1}$ coincide
with the the typical and disorder averaged correlation lengths of the
system, respectively.

As we discuss briefly in Section \ref{sec:gen}, the finite--size
estimates of Theorem \ref{th:fss} and the bounds of Theorem
\ref{th:bounds} have a very natural generalization to the case of
random copolymers at a selective interface between two solvents
\cite{cf:Monthus,cf:BdH,cf:G}, which also show a
localization/delocalization transition.  In this case, the relevant
order parameter is not the contact fraction but the fraction of
monomers in the unfavorable solvent.

\section{Random pinning models}
\label{sec:model}
Let $S=\{S_n\}_{n=0,1,\ldots}$ be a time--homogeneous process with law
$\bP$, taking values in some set $\Sigma$ and such that
$S_0=0\in\Sigma$. We will be especially interested in the returns to
zero of $S$: we let $\tau_0=0$ and, for $i\ge 1$,
$\tau_i=\inf\{j>\tau_{i-1}: S_j=0\}$. If $\tau_i=\infty$,
then by convention $\tau_{i+1}=\infty$.  The only assumptions
we  make on $\bP$ is that 
$\{\tau_i-\tau_{i-1}\}_{i=1,2,\ldots}$ is a sequence of IID random
variables taking values in $\N \cup \{\infty\}$ and 
that, defining $K(n):=\bP(\tau_1=n)$, there exists $s\in \N$ such that 
\begin{eqnarray}
\label{eq:RW}
K(s n) =\frac{L(n)}{n^\alpha}, 
\end{eqnarray}
and $K(n)=0$ if $n\notin s\N$, for some $1\le \alpha<\infty$ and a
function $L(\cdot)$ varying slowly at infinity, i.e., a positive
function such that $\lim_{x\to\infty} L(x r)/L(x)=1$ for every $r>0$
\cite{cf:Feller2}. An example of slowly varying function
is 
 $r \mapsto (\log (r+1))^b$, for  $b\in \R$, 
 but also $r \mapsto \exp( (\log (r+1))^b)$,  for $b <1$, 
as well as any positive function for which $\lim_{r\to \infty} L(r)>0$. 

On the {\em defect line} $S\equiv 0$ are placed random charges
$\go=\{\go_n\}_{n=1,2,\ldots}$ which we assume to be IID {\em bounded}
random variables with law $\bbP$. We will assume that $\bbE\,[
\go_1]=0$ and $\bbE\,[ \go_1^2]=1$ (which, as will be clear from \eqref{eq:H}
below, implies no loss of generality).  The Hamiltonian
describing the interaction between the polymer and the defect line
depends on two parameters, $\beta\ge0$ (playing the role of the
strength of the disorder) and $h\in\R$ (where $-h$ represents the
average energetic gain of a polymer--line contact):
\begin{eqnarray}
  \label{eq:H}
\mathcal H_{N,\go}^{\beta,h}(S)=\sum_{n=1}^{N}\left(\beta\go_{n} - h\right)\ind_{\{S_n=0\}}.
\end{eqnarray}
The corresponding Boltzmann distribution is
\begin{equation}
\label{eq:Boltzmann}
\frac {\dd \bP^{\beta,h} _{N, \go}} {\dd \bP} (S)
\, = \,
\frac {e^{\mathcal H_{N,\go}^{\beta,h}(S)} }
{ Z_{N,\go}^{\beta,h}}
 \ind_{\left\{ S_N=0\right\}}
\end{equation}
and, of course, the partition function is given by
\begin{equation}
\label{eq:Z}
  Z_{N,\go}^{\beta,h}=\bE\left(e^{\mathcal H_{N,\go}^{\beta,h}(S)} \ind_{\left\{ S_N=0\right\}}\right).
\end{equation}
Here and in the following, we assume that $N\in s\N$,
even when not explicitly stated.

As Equation \eqref{eq:Boltzmann} shows, the polymer tends to touch the
defect line at points where $\beta\go_n-h>0$ and to avoid it in the
opposite situation. Note that there is a competition between an
energetic effect (trying to touch as many favorable points as possible along the
line) and an entropic one (trajectories which stay close to the line
are much less numerous than those which wander away).  Therefore, it is
quite intuitive (and actually well known) that a (de)localization
transition takes place when the strength of the polymer--line
interaction is varied. This will be discussed below.

\begin{rem}\rm
  We restrict to bounded disorder variables $\go_n$ just for
simplicity of exposition. The results below can be extended to more
general situations but we will not pursue this line. Let us just
mention that all the results of this paper hold also in the Gaussian
case $\go_1=\mathcal N(0,1)$.  In more general cases of continuous, unbounded
disorder variables, 
a sufficient condition for the results to hold
is that the sub--Gaussian concentration
inequality \eqref{eq:eq_Lip} is satisfied by $\bbP$
and that a certain condition on the smoothness of the density of
$\go_1$ with respect to the Lebesgue measure on $\R$ holds
(cf. \cite{cf:GT05}, condition {\bf C2}).  A discussion of the
relevance of concentration of measure inequalities in pinning and
copolymer models can be found in \cite{cf:GT}.
\end{rem}
\begin{rem}
\rm
Note that only  the model with endpoint $S_N$ pinned to zero is being
considered, cf. Eq. \eqref{eq:Boltzmann}. This is just for simplicity of 
exposition, since this way one has for $M<N$
\begin{equation}
 \label{eq:super} 
\log \Zno\ge \log Z_{M,\go}^{\beta,h}+\log Z_{N-M,\theta^{M}\go}^{\beta,h}
\end{equation}
($\theta$ is the left shift: $\theta \go_n=\go_{n+1}$), a
property we will use several times in the proofs of Section
\ref{sec:proofs}.  By the way, note that \eqref{eq:super} implies that
the sequence $\{\bbE \log Z_{N,\go}^{\beta,h}\}_N$ is super--additive
in $N$.  One could also leave the endpoint free: in this case, in the
r.h.s. of Eq. \eqref{eq:super} error terms of order $\log N$ would
appear (cf. e.g. \cite[Remark 1.1]{cf:GT05}). As a consequence, in the
proof of the theorems one would have to keep track of harmless but
annoying logarithmic error terms.
\end{rem}

\begin{rem}\rm
To make condition \eqref{eq:RW} more explicit note that, for instance,
if $\{S_n\}_n$ is the SRW (simple random walk) on $\Sigma=\Z^d$, then
\eqref{eq:RW} holds with $s=2$ and $\alpha=3/2$ for $d=1$ and
$\alpha=d/2$ for $d\ge 2$.  The Poland--Scheraga model of DNA
denaturation also fits into our framework; in this case, the
physically relevant value of $\alpha$ is around $2.11$ \cite{cf:KMP}.
For the Poland--Scheraga model, the contact fraction defined in
Eq. \eqref{eq:ellN} below corresponds to the fraction of bound base
pairs.
\end{rem}

As it is well known the infinite--volume free energy, i.e. the limit
\begin{eqnarray}
  \tf(\beta,h)=\lim_{N\to\infty}\frac1N \log Z^{\beta,h}_{N,\go}
\end{eqnarray}
exists, is almost--surely independent of $\go$ and satisfies
$\tf(\beta,h)\ge0$ (cf. e.g. \cite{cf:AS}, \cite{cf:G}, 
but proofs of these facts
have appeared several times in the literature. The non--negativity of
$\tf$ is proven by simply restricting the average in \eqref{eq:Z} to
the configurations which do not touch zero between sites $0$ and $N$,
and using Eq. \eqref{eq:RW}.) One decomposes the phase diagram
$(\beta,h)$ into depinned (or delocalized) and pinned (or localized)
phases, $\cD$ and $\cL$, defined as $\cD=\{(\beta,h):\tf(\beta,h)=0\}$
and $\cL=\{(\beta,h):\tf(\beta,h)>0\}$, separated by a critical line
$h_c(\beta)=\inf\{h:\tf(\beta,h)=0\}$.  Various properties of the
critical curve are known \cite{cf:AS} \cite{cf:G}: in particular,
under our assumptions one has that, for every $0<\beta<\infty$,
\begin{equation}
\label{eq:curva}
h_c(0)=\log (1-\bP(\tau_1=\infty))<h_c(\beta)<\infty.  
\end{equation}
Note that $h_c(0)\le 0$, and $h_c(0)<0$ iff $S$ is transient.
Moreover, $h_c(\cdot)$ is a convex increasing function, as follows
easily from the convexity of $\tf$ with respect to its arguments and
from \eqref{eq:curva}.

The order parameter associated to the (de)localization transition is
the {\em contact fraction}, defined as
\begin{eqnarray}
\label{eq:ellN}
\ell_N:=\frac{\cf}N:=\frac{\vert\{1\le n\le N: S_n=0\} \vert}N.  
\end{eqnarray}
Since  $\tf$ is clearly convex as a function of $h$, and since it is 
differentiable in 
$h$ for every $h<h_c(\beta)$ (as was proven in  \cite{cf:GTloc}), from
the definitions of $\cL,\cD$ it follows that, $\bbP(\dd \go)$--a.s.,
\begin{equation}
 \lim_{N\to\infty}\bE^{\beta,h}_{N,\go}(\ell_N)=-\partial_h\tf(\beta,h)>0\;\;\; \mbox{if}\;\;\; h<h_c(\beta) 
\end{equation}
while
\begin{equation}
\label{eq:cf_D}
    \lim_{N\to\infty}\bE^{\beta,h}_{N,\go}(\ell_N)=0 \;\;\; \mbox{if}\;\;\; h>h_c(\beta).
\end{equation}
However, much more than \eqref{eq:cf_D} is true: indeed, in \cite{cf:GT} it was proven that, for  $h>h_c(\beta)$,
\begin{equation}
  \bbE\, \bP^{\beta,h}_{N,\go}(\cf \ge m)\le e^{-d_1 m}
\end{equation}
if $m\ge d_2 \log N$, for some constants
$0<d_1(\beta,h),d_2(\beta,h)<\infty$. In other words, 
 the number of 
contacts with the defect line 
grows, typically, linearly with $N$ for $h<h_c(\beta)$ and
at most logarithmically in $N$ for $h>h_c(\beta)$. Finally, in
\cite{cf:GT05}--\cite{cf:GTlett} it was proven that
$\partial_h\tf(\beta,h)$ vanishes continuously for $h\uparrow
h_c(\beta)$ if $\beta>0$, which implies that, $\bbP(\dd \go)$--a.s.,
\begin{equation}
  \lim_{N\to\infty}\bE^{\beta,h_c(\beta)}_{N,\go}(\ell_N)=0.
\end{equation}
In view of these facts, it is very natural to ask what is the typical
size of the contact fraction for finite $N$, {\em at} the critical
point or very close to it. This question will be addressed in the next
section.

\section{Main results}

\subsection{Finite--size estimates on the contact fraction}
Since we are interested in the finite--size
scaling behavior of the system in a window around the critical point,
shrinking to zero with the system size, we allow in general $h$
to depend on $N$, and write explicitly $h=h_N$.
\begin{theorem}
\label{th:fss}
  Let  $\beta>0$ and $1\le \alpha<\infty$. Assume that 
  \begin{eqnarray}
    \label{eq:scaling}
\lim_{N\to\infty} N^{t} (h_N-h_c(\beta))=b     \in \R
  \end{eqnarray}
for some $t\ge0$. Then, 
  \begin{enumerate}
  \item If $t\ge 1/3$, then for $c$ sufficiently large
  \begin{eqnarray}
    \label{eq:main}
    \lim_{N\to\infty}\bbE \, \bP^{\beta,h_N}_{N,\omega}\left(\mathcal N_N\ge  c N^{2/3}\log N\right)=0
  \end{eqnarray}
\item If $t<1/3$ and $b>0$, then for $c$ sufficiently large
\begin{eqnarray}
    \label{eq:main2}
    \lim_{N\to\infty}\bbE \, \bP^{\beta,h_N}_{N,\omega}\left(\mathcal
    N_N\ge c N^{2t}\log N\right)=0.
  \end{eqnarray}
\item If $t<1/3$ and $b<0$, then for $c$ sufficiently large
\begin{eqnarray}
    \label{eq:main3}
    \lim_{N\to\infty}\bbE \, \bP^{\beta,h_N}_{N,\omega}\left(\mathcal N_N\ge  c N^{1-t}\right)=0.
  \end{eqnarray}
  \end{enumerate}
\end{theorem}
It is understood that the constant $c$ above can depend on $\beta$,
$\alpha$ and $b$.  Note that, for $t=0$ and $b>0$, one finds back
the known estimates on the contact fraction valid in the interior of $\cD$ \cite{cf:GT}.
\begin{rem} \rm
The estimates of Theorem \ref{th:fss} need not be optimal, in
general. Indeed, as will be clear in Section \ref{sec:proofs}, our
proof is based on the fact that $\tf$ vanishes at least quadratically
when the critical line is approached from the localized region and $\beta>0$ \cite{cf:GT05}:
\begin{equation}
\label{eq:smooth}
  \tf(\beta,h)\le \alpha c_1(\beta)(h_c(\beta)-h)^2
\end{equation}
for some constant $0<c_1(\beta)<\infty$, if $h<h_c(\beta)$.
 On the other hand it
is  quite reasonable, and actually expected in the physics
literature, that the transition is smoother in various situations, for
instance if $\alpha\le 3/2$ and $\beta$ small. 
Following the proof of Theorem \ref{th:fss} 
in Section \ref{sec:proofs} it is not difficult  to
realize (cf. Remark \ref{rem:ipo} below) that, if one assumes 
\begin{eqnarray}
  \label{eq:supponiamo}
  \tf(\beta,h)\le c_\tf(\beta,\alpha) (h_c(\beta)-h)^k
\end{eqnarray}
for every $h<h_c(\beta)$ then, for instance, 
\begin{eqnarray}
  \label{eq:ipotetica}
  \lim_{N\to\infty}\bbE \, \bP^{\beta,h_c(\beta)}_{N,\omega}\left(\mathcal N_N\ge  c N^{2/(k+1)}\log N\right)=0,
\end{eqnarray}
for $c$ sufficiently large. 
If $k>2$, this would clearly improve the upper bound on the contact fraction at
the critical point given by Theorem \ref{th:fss}.
Estimates \eqref{eq:main}--\eqref{eq:main3} could also be similarly
improved for all values of $t$ and $b$. Unfortunately, up to now there
are no known cases where one can prove an estimate like
\eqref{eq:supponiamo}, with $k>2$, for non--zero values of $\beta$.
\end{rem}

\subsection{$\mu$ versus $\rm F$: an inequality for critical exponents}
\label{sec:versus}
In Refs. \cite{cf:AZ} and \cite{cf:GTloc}, the quantity
\begin{eqnarray}
  \label{eq:mu}
  \mu(\beta,h)=-\lim_{N\to\infty}\frac1N \log\bbE \left[\frac1 {Z^{\beta,h}_{N,\go}}\right]
\end{eqnarray}
was introduced. As it was proved there, in the localized phase $\mu$
is strictly positive and related to maximal excursions of the polymer
from the defect line: indeed, for the polymer of length $N$ the
maximal distance between two successive returns to zero of $S$ is
typically $(\log N)/\mu(\beta,h)$.  When $h$ approaches $h_c(\beta)$
from below, $\mu$ tends to zero and therefore the length of the
maximal excursion diverges, on the scale $\log N$. More precisely, the
following bounds were proven in \cite{cf:GTloc}: for every $\beta>0$
there exists $0<c_2(\beta)<\infty$ such that
\begin{eqnarray}
  \label{eq:bound_mu_0}
c_2(\beta)\tf(\beta,h)^2<\mu(\beta,h)<\tf(\beta,h),
\end{eqnarray}
where the lower bound holds, say, for  $0<h_c(\beta)-h\le 1$.
Our next result significantly improves the lower bound in \eqref{eq:bound_mu_0}:
\begin{theorem}
\label{th:bounds}
For every $\beta>0$ there exists $0<c_3(\beta)<\infty$ such that
  \begin{eqnarray}
\label{eq:bound_mu}
  0< -c_3(\beta)\frac{\tf(\beta,h)^2}{\partial_h\tf(\beta,h)} <\mu(\beta,h)
  \end{eqnarray}
if $0<h_c(\beta)-h\le 1$.
\end{theorem}
\begin{rem}\rm 
In order to give a more readable form to these bounds assume that, for $\beta>0$ and $h< h_c(\beta)$,
\begin{eqnarray}
  \label{eq:supponiamo2}
\tf(\beta,h)=c_\tf\left(\beta,(h_c(\beta)-h)^{-1}\right)(h_c(\beta)-h)^{\nu_\tf}
\end{eqnarray}
and 
\begin{eqnarray}
  \label{eq:supponiamo3}
\mu(\beta,h)= c_\mu\left(\beta,(h_c(\beta)-h)^{-1}\right)(h_c(\beta)-h)^{\nu_\mu}
\end{eqnarray}
for some functions $c_\tf(\beta,x),c_\mu(\beta,x)$ slowly varying in $x$
for $x\to\infty$ 
(of course $\nu_\tf,\nu_\mu\ge 2$, as a consequence of Eq. 
\eqref{eq:smooth} and of the upper bound in \eqref{eq:bound_mu_0}; in principle, $\nu_\tf,\nu_\mu$ can depend on 
$\beta$).
Then, 
recalling the definition of slow variation and the fact that $\tf$ is convex
in $h$, one realizes that Eq. 
 \eqref{eq:bound_mu_0} implies
\begin{eqnarray}
\label{eq:bound_nu_0}
(2\le )\nu_\tf\le \nu_\mu\le 2\nu_\tf.
\end{eqnarray}
while from \eqref{eq:bound_mu} follows that
\begin{eqnarray}
\label{eq:bound_nu}
\nu_\mu\le \nu_\tf+1.
\end{eqnarray}
\end{rem}

\subsection{Typical and average correlation lengths for a $(1+1)$--dimensional
wetting model}

\label{sec:xi}

Beyond giving informations about the divergence of the longest
excursion close to (but below) the critical line, bounds like
\eqref{eq:bound_mu} involving $\mu$ and $\tf$ are of interest because
it is rather natural to expect that  
$\mu^{-1}$ (respectively
$\tf^{-1}$)  
has the same divergence, for $h$ approaching $h_c(\beta)$ from the
localized phase, as 
 the average (respectively typical) correlation length of
the system.
Our next result, Theorem
\ref{th:correlazioni}, makes this conjecture precise at least in a
specific model of $(1+1)$--dimensional wetting. 

 Recall that in
\cite[Theorem 2.2]{cf:GTloc} it was proven that, for every bounded
local observable $A$ (i.e., bounded function which depends  on
$S_j$ only for $j$ in a finite subset of $\N$), the infinite--volume limit
\begin{eqnarray}
  \bE^{\beta,h}_{\infty,\go}(A)=\lim_{N\to\infty} \bE^{\beta,h}_{N,\go}(A)
\end{eqnarray}
exists $\bbP(\dd \go)$--almost surely, if $(\beta,h)\in\cL$. Moreover, in $\cL$
truncated correlation functions decay exponentially fast with
distance. In fact, for every bounded local observables $A,B$ define
the local observable $B_k$ as $B_k(S)=B(\theta^k S)$, where $\theta$
is the left shift, $\theta S_n=S_{n+1}$. Then, there exist a constant
$0<c_{A,B}(\beta,h)<\infty$, an almost surely finite random variable
$C_{A,B}(\go,\beta,h)$ and a constant $d(\beta,h)>0$ such that
\cite{cf:GTloc}, in $\cL$,
\begin{eqnarray}
 \label{eq:exp_av}  
\bbE \left|\bE_{\infty,\go}^{\beta,h}(A B_k)-\bE_{\infty,\go}^{\beta,h}(A)
\bE_{\infty,\go}^{\beta,h}(B_k)\right|\le c_{A,B}e^{-d(\beta,h)k}
\end{eqnarray}
and
\begin{eqnarray}
 \label{eq:exp_as}  
\left|\bE_{\infty,\go}^{\beta,h}(A B_k)-\bE_{\infty,\go}^{\beta,h}(A)
\bE_{\infty,\go}^{\beta,h}(B_k)\right|
\le C_{A,B}(\go)e^{-d(\beta,h)k}.
\end{eqnarray}
However, in \cite{cf:GTloc} the $(\beta,h)$ dependence of the constant
$d(\beta,h)$ was not tracked, and lower bounds complementary to Eqs.
\eqref{eq:exp_av}, \eqref{eq:exp_as} were not obtained. It turns out
that this gap can be filled, at least in the case of a rather natural
$(1+1)$--dimensional wetting model we define now.  This model still
belongs to the class described by the Boltzmann distribution
\eqref{eq:Boltzmann} but, in addition to the basic assumptions of
Section \ref{sec:model}, we require that the state space of the
process $S$ is $\Sigma=\Z^+$ (i.e., there is an impenetrable wall
which prevents $S_n<0$) and that actually $S$ is the SRW 
with increments $S_{i+1}-S_i=\pm1 $, conditioned to be
non--negative (the condition $|S_i-S_{i-1}|=1$ could be somewhat
relaxed in the theorem below, at the price of some further technical
work. We will not pursue this line).  Note that in this case
\eqref{eq:RW} holds with $\alpha=3/2$ and $s=2$.  This model has a
natural interpretation as a {\em $(1+1)$--dimensional wetting model of
  a disordered substrate} \cite{cf:FLNO,cf:DHV,cf:AS}.  The defect
line represents a wall with impurities, and $S$ the interface between
two coexisting phases (say, liquid below the interface and vapor
above).  When $h<0$ the underlying homogeneous substrate repels the
liquid phase, and vice versa for $h>0$.  $\cL$ corresponds then to the
{\em dry phase} (microscopic liquid layer at the wall) and $\cD$ to
the {\em wet phase} (macroscopic layer).

Then, one has:
\begin{theorem}
\label{th:correlazioni}
For the wetting model just introduced, the following holds: for every
$\beta\ge0$ and $h<h_c(\beta)$,
\begin{eqnarray}
  \label{eq:corr_ave}
-\lim_{k\to\infty}\frac1k \log \bbE
\left(\bP_{\infty,\go}^{\beta,h}(S_\ell=S_{\ell+k}=0)-
\bP^{\beta,h}_{\infty,\go}(S_\ell=0)
\bP^{\beta,h}_{\infty,\go}(S_{\ell+k}=0)\right)=\mu(\beta,h)
\end{eqnarray}
and, $\bbP(\dd\go)$--a.s., 
\begin{eqnarray}
  \label{eq:corr_typ}
-\lim_{k\to\infty}\frac1k \log
\left(\bP^{\beta,h}_{\infty,\go}(S_\ell=S_{\ell+k}=0)-\bP_{\infty,\go}^{\beta,h}(S_\ell=0)
\bP^{\beta,h}_{\infty,\go}(S_{\ell+k}=0)\right)=\tf(\beta,h).
\end{eqnarray}
\end{theorem}
Here it is understood that $\ell,k,N\in 2\N$, due to the periodicity of
the simple random walk. 

\begin{rem}\rm
It would be extremely interesting, especially in view
of Theorem \ref{th:correlazioni}, to fill the gap between the
 upper bound in \eqref{eq:bound_mu_0} and the lower bound
 \eqref{eq:bound_mu} (or equivalently, between \eqref{eq:bound_nu_0}
 and \eqref{eq:bound_nu}). In the case of the $(1+1)$--dimensional
 wetting model with $\pm1$ increments, this would answer the question
  whether typical and average correlation lengths have the same
 critical behavior close to the depinning transition, or if their
 divergence is governed by different critical exponents, as it happens
 for instance in the disordered Ising spin chain with random
 transverse field of Ref. \cite{cf:fisher}.
\end{rem}

\section{Generalization to copolymers at a selective interface}

\label{sec:gen}

In this Section we sketch briefly how Theorems \ref{th:fss} and
\ref{th:bounds} can be extended to the model of {\em random copolymer
  at a selective interface} \cite{cf:BdH,cf:G,cf:Monthus}. We refer
for instance to \cite{cf:SW,cf:G} for physical motivations of this
model.  In this case, the state space  of $S$ is $\Sigma\equiv\Z$ and, in
addition to time homogeneity of $S$ and to the IID property of
the sequence $\{\tau_i-\tau_{i-1}\}_i$, one assumes that
$(S_{i+1}-S_i)\in\{-1,0,+1\}$ and that $\bP$ is invariant under the
transformation $S\to -S$. The Hamiltonian \eqref{eq:H} is replaced by
\begin{equation}
\label{eq:Hc}
  \hat {\mathcal H}_{N,\go}^{\beta,h}(S)=\sum_{n=1}^N (\beta \go_n-h)\ind_{\{S_n<0\}}
\end{equation}
where, without loss of generality in view of the symmetry of $\bP$, we
can assume that $h\ge0$.  The variables $\{\go_n\}_n$ are IID centered
and satisfy the same boundedness assumption as in Section
\ref{sec:model}.  The Boltzmann distribution and the partition
function $\hat Z_{N,\go}^{\beta,h}$ are defined as in Eqs.
\eqref{eq:Boltzmann}, \eqref{eq:Z}, provided that $\Hno$ is replaced
by $\hat {\mathcal H}_{N,\go}^{\beta,h}$.  One should imagine the
model as describing a polymer $S$ in proximity of the interface
($S\equiv 0$) between two solvents A and B, placed in the half--planes
$S>0$ and $S< 0$, respectively.  Note that $S_n$ has the tendency to
be in A whenever $\beta \go_n-h<0$ and in B if $\beta \go_n-h>0$. Note
also that, if $h>0$, for a typical disorder realization the polymer
has a net preference to be in A, which will be called the {\em
  favorable solvent}.

Again, it is known \cite{cf:BdH} that the infinite--volume free energy
$\hat\tf(\beta,h)=\lim_N(1/N)\log \hat Z_{N,\go}^{\beta,h}$ exists, is
almost surely independent of $\go$ and non--negative, so that one
can define the localized and delocalized phases, $\hat\cL$ and $\hat\cD$,
is analogy to Section \ref{sec:model}. 
Upper \cite{cf:BdH} and lower
\cite{cf:BG} bounds are known for the critical curve ${\hat h}_c(\beta)
=\inf\{h:\hat\tf(\beta,h)=0\}$ but,
on the basis of careful numerical simulations plus concentration of measure
considerations, none of them is believed to be optimal in general
\cite{cf:CGG}.  In contrast with the case of the pinning models of
Section \ref{sec:model}, for the copolymer the order parameter
associated to the localization/delocalization transition is the {\em
fraction of monomers in the unfavorable solvent}:
\begin{equation}
  \hat \ell_N:=\frac{\hat {\mathcal N}_N}N:=\frac{|\{1\le n\le N\}: S_n<0\}|}N.
\end{equation}
This is rather intuitive since, comparing definitions \eqref{eq:H} and
\eqref{eq:Hc}, one notices that the role of $\ind_{\{S_n=0\}}$ is now
played by $\ind_{\{S_n<0\}}$.  Like for the contact fraction in
pinning models, various estimates on the order parameter are known:
$\hat\ell_N$ is of order $1$ in $\hat\cL$, at most of order $(\log
N)/N$ in the interior of $\hat\cD$ \cite{cf:GT} and $o(1)$ for
$N\to\infty$ at the critical line \cite{cf:GT05}. The methods we
introduce in the present paper allow to make the last statement
sharper: indeed, Theorem \ref{th:fss} holds unchanged also for the
copolymer model, provided that $\cf$ is replaced by $\hat {\mathcal
  N}_N$.  In particular, therefore, $\hat \ell_N$ is at most of order
$N^{-1/3}\log N$ at the critical point.

Theorem \ref{th:bounds} also admits a natural extension to the
copolymer case: if $\hat \mu(\beta,h)$ is defined as in \eqref{eq:mu},
with $\Zno$ replaced by $\hat Z_{N,\go}^{\beta,h}$, then again
Eq. \eqref{eq:bound_mu} holds with $\tf,\mu$ replaced by $\hat \tf,\hat
\mu$.

In order to avoid a useless duplication of the proofs of Theorems
\ref{th:fss} and \ref{th:bounds}, in Section \ref{sec:proofs} we will
consider only the case of pinning models and we will not give details
for the copolymer case: as it was also the case in Refs.
\cite{cf:GT}--\cite{cf:GTloc}, it is easy to realize that the two
models can be treated analogously, if the correct order parameter is
used in each case. Just to give an example, Eq. \eqref{eq:conc_gen}
below holds also for the copolymer, if $\cf$ is replaced by $\hat
{\mathcal N}_N$, as was proven in \cite[Lemma 2.1]{cf:GT}.

\section{Proof of the results}

\label{sec:proofs}

Given a set $\Omega$ of polymer configurations, measurable with
respect to $\bP$, it is convenient to set
\begin{eqnarray}
\label{eq:Zvinc}
  Z_{N,\go}^{\beta,h}(\Omega):=\bE \left ( e^{{\mathcal
H}_{N,\go}^{\beta,h}(S)}\ind_{\{S\in \Omega\}} \ind_{\{S_N=0\}}
\right).
\end{eqnarray}

Our basic technical tool is the following classical concentration
inequality \cite{cf:Ledoux_libro}: if $\go=\{\go_n\}_n$ is a sequence
of IID bounded random variables with law $\bbP$, there exist constants
$0<C_1,C_2<\infty$ such that, for every convex Lipschitz function
$f:\R^n\to \R$, one has:
\begin{eqnarray}
  \label{eq:eq_Lip}
  \bbP\left(\left|f(\go_1,\ldots,\go_n)-\bbE f(\go_1,\ldots,\go_n)\right|\ge t\right)\le C_1 \exp
\left(-\frac{C_2 t^2}{||f||^2_{Lip}}\right)
\end{eqnarray}
for every $t>0$, where $||f||_{Lip}$ is the Lipschitz norm of $f$ with
respect to the Euclidean norm in $\R^n$, i.e., the smallest $M\ge0$ 
such that
\begin{equation}
  \suptwo{x, y\in \R^n}{x\ne y}\frac{\left|f(x)-f(y)\right|}
{\left[\sum_{i=1}^n(x_i-y_i)^2\right]^{1/2}}\le M.
\end{equation}
The way we will employ this inequality is by noting that $(1/N) \log
Z_{N,\go}^{\beta,h}$, considered as a function of
$\go_1,\ldots,\go_N$, is convex and has a Lipschitz constant at most
$\beta/\sqrt N$.  More generally, one has the following \cite[Lemma
2.1]{cf:GT}: let $\Omega_m$ be a set of polymer trajectories such that
$\mathcal N_N\le m$ for every $S\in\Omega_m$. Then,
\begin{eqnarray}
  \label{eq:conc_gen}
\bbE\left(\left|\frac1N \log Z^{\beta,h}_{N,\go}(\Omega_m)-
\frac1N \bbE\log Z^{\beta,h}_{N,\go}(\Omega_m)\right|\ge t\right)\le
C_1 \exp
\left(-C_2\frac{N^2 t^2}{\beta^2 m}\right).
\end{eqnarray}
This is simply proven by noting that $(1/N) \log
Z^{\beta,h}_{N,\go}(\Omega_m)$ has a Lipschitz constant at most
$\beta\sqrt m /N$.

\subsection{ Proof of Theorem \ref{th:fss}}

For $m\in \N\cup\{0\}$, consider the restricted partition function
\begin{eqnarray}
  \label{eq:z_vinc}
  Z_{N,\go}^{\beta,h}(\mathcal N_N=m)
\end{eqnarray}
where the number of contacts with the line, $\mathcal N_N$, is
 constrained to $m$.  Thanks to the fact that the differences
 $\tau_i-\tau_{i-1}$ between successive return times to zero of $S$
 are independent under the law $\bP$, one has
\begin{eqnarray}
  \label{eq:superadd}
  \frac1N \bbE \log Z_{N,\go}^{\beta,h}(\mathcal N_N=m)\le
\lim_{k\to\infty} \frac1{kN}\bbE \log Z_{kN,\go}^{\beta,h} (\mathcal
N_{kN}=k m) \le \phi\left(\beta,\frac mN\right)-h\frac mN,
\end{eqnarray}
where
\begin{eqnarray}
  \label{eq:phi}
  \phi(\beta,x)=\lim_{\varepsilon\searrow0}\lim_{N\to\infty}\frac1N
\bbE \log Z^{\beta,0}_{N,\go}\left(
\ell_N\in [x-\varepsilon,
x+\varepsilon]\right).
\end{eqnarray}
The limits in \eqref{eq:phi} exist for monotonicity reasons.
In \cite{cf:GT05} it was proven that, under some assumptions on $\bbP$
(assumptions which are satisfied, in particular, in the case of bounded
random variables we are considering here), one has for $\beta>0$ 
\begin{eqnarray}
  \label{eq:stimaphi}
  \phi(\beta,x)\le -\frac{c_4(\beta)}\alpha x^2+h_c(\beta)x,
\end{eqnarray}
for some constant $0<c_4(\beta)<\infty$ depending only on the law $\bbP$.
\begin{rem}\rm
\label{rem:ipo}
Equation \eqref{eq:stimaphi} follows simply from Eq. \eqref{eq:smooth} and
from 
the fact that, as was proven in \cite{cf:GT05},
 $\tf$ is related to the function $\phi$ of Eq. \eqref{eq:phi} via the Legendre
transformation
\begin{eqnarray}
\label{eq:legendre}
  \tf(\beta,h)=\sup_{x\in[0,1]}\left(\phi(\beta,x)-hx\right).
\end{eqnarray}
(Actually, in \cite{cf:GT05} the reverse path was followed: first
\eqref{eq:stimaphi} was proven, and then \eqref{eq:smooth} was
deduced).  If one could prove Eq. \eqref{eq:supponiamo} with $k>2$,
\eqref{eq:stimaphi} would be immediately improved into
\begin{eqnarray}
  \label{eq:stimaphi_ipo}
  \phi(\beta,x)\le -\tilde c_{\tf}(\beta,\alpha)x^{k/(k-1)}+h_c(\beta)x
\end{eqnarray}
for some $0<\tilde c_{\tf}(\beta,\alpha)<\infty$.
\end{rem}
\medskip

Equation \eqref{eq:stimaphi}, together with \eqref{eq:superadd}, implies that for every $N\in s\N$,  $m\in \N\cup\{0\}$
\begin{eqnarray}
  \label{eq:sothat}
  \frac1N \bbE \log Z_{N,\go}^{\beta,h_N}(\mathcal N_N=m)\le -
  \frac{c_4(\beta)}\alpha \left(\frac mN\right)^2-(h_N-h_c(\beta))\frac mN.
\end{eqnarray}

Let us consider first the case $b>0$, $t<1/3$. Then,
for $N$ sufficiently large one has, uniformly in $m$,
\begin{eqnarray}
  \label{eq:sothat2}
  \frac1N \bbE \log Z_{N,\go}^{\beta,h_N}(\mathcal N_N=m)\le -\frac b2N^{-t}\frac mN.
\end{eqnarray}
We let $E_1$ be the event
\begin{eqnarray}
  \label{eq:eventoE}
E_1=\left\{ \text{there exists}\; m\ge c N^{2t}\log N \;\text{such that}\; 
\frac1N  \log Z_{N,\go}^{\beta,h_N}(\mathcal N_N=m)\ge -\frac b4 N^{-t}\frac mN
\right\}.
\end{eqnarray}
To estimate the probability of $E_1$, we employ Eq. \eqref{eq:conc_gen}
and we find
\begin{eqnarray}
  \label{eq:probE1}
  \bbP[E_1]\le C_1\sum_{m\ge c  N^{2t}\log N} e^{-C_2\frac{ b^2 m N^{-2t}}{16\beta^2}}
\end{eqnarray}
which decays to zero for $N\to\infty$, if $c$ is large enough.
On the complementary of the event $E_1$, one the other hand, one has
\begin{eqnarray}
\label{eq:complE1}
  \bP_{N,\go}^{\beta,h_N}(\mathcal N_N\ge c  N^{2t}\log N)&=&
\frac{\sum_{m\ge c N^{2t}\log N}Z_{N,\go}^{\beta,h_N}(\mathcal N_N=m)
}{Z_{N,\go}^{\beta,h_N}}\\
\nonumber&
\le& c_5 N^{2\alpha} \sum_{m\ge c N^{2t}\log N } e^{- \frac{b m N^{-t}}4}
\end{eqnarray}
which also decays to zero. In \eqref{eq:complE1} we used the obvious bound
\begin{equation}
\label{eq:nome}
Z_{N,\go}^{\beta,h}\ge Z_{N,\go}^{\beta,h}\left(\{S_n\ne 0\;\mbox{for every} \;n<N\}\right)=K(N)e^{\beta\go_N-h}\ge (c_5)^{-1}
 N^{-2\alpha},
\end{equation}
cf. Eq. \eqref{eq:RW} and the definition of slowly varying function.
Equations \eqref{eq:probE1} and \eqref{eq:complE1} together imply \eqref{eq:main2}.

Next, consider the case $t\ge1/3$.  It is immediate to check that, for
$N$ sufficiently large and $m\ge c N^{2/3}\log N$, the r.h.s. of
Eq. \eqref{eq:sothat} is smaller than
$$
-\frac{c_4(\beta)}{2\alpha}\left(\frac mN\right)^2.
$$
Then, one defines 
\begin{eqnarray}
  \label{eq:eventoE2}
E_2=\left\{ \text{there exists}\; m\ge c N^{2/3}\log N \;\text{such that}\; 
\frac1N  \log Z_{N,\go}^{\beta,h_N}(\mathcal N_N=m)\ge -\frac{c_4(\beta)}{4\alpha}\left(\frac mN\right)^2
\right\}
\end{eqnarray}
and notes that, in analogy with Eqs. \eqref{eq:probE1} and \eqref{eq:complE1}, 
\begin{eqnarray}
  \label{eq:probE2}
  \bbP[E_2]\le C_1\sum_{m\ge c  N^{2/3}\log N} e^{-C_2\frac{ c_4(\beta)^2  m^3 }{16\alpha^2N^2\beta^2}}
\end{eqnarray}
while, on the complementary of the event $E_2$, 
\begin{eqnarray}
\label{eq:complE2}
  \bP_{N,\go}^{\beta,h_N}(\mathcal N_N\ge c  N^{2/3}\log N)
\le c_5 N^{2\alpha} \sum_{m\ge c N^{2/3}\log N } e^{- \frac{c_4(\beta) m^2}{4N}},
\end{eqnarray}
which together imply \eqref{eq:main}, for $c$ large.

Finally, the case $b<0$ and $t< 1/3$.  One realizes easily that, for
$N,c$ sufficiently large and $m\ge c N^{1-t}$, the r.h.s. of
Eq. \eqref{eq:sothat} is smaller than
$$
-|b| N^{-t}\frac mN.
$$
Then, one defines 
\begin{eqnarray}
  \label{eq:eventoE3}
E_3=\left\{ \exists m\ge c N^{1-t} \;\text{such that}\; \frac1N
\log Z_{N,\go}^{\beta,h_N}(\mathcal N_N=m)\ge -\frac{| b|}2 N^{-t}\frac mN
\right\}
\end{eqnarray}
and notes that
\begin{eqnarray}
  \label{eq:probE3}
  \bbP[E_3]\le C_1\sum_{m\ge   c N^{1-t}} e^{-C_2 \frac {b^2}{4\beta^2}N^{-2t}m}
\end{eqnarray}
which decays to zero for $N\to\infty$ since $t< 1/3$
while, on the complementary of the event $E_3$, 
\begin{eqnarray}
\label{eq:complE3}
  \bP_{N,\go}^{\beta,h_N}(\mathcal N_N\ge c N^{1-t})
\le c_5 N^{2\alpha} \sum_{m\ge  c N^{1-t} } e^{- \frac {|b|}2 N^{-t}m}.
\end{eqnarray}
 Equation \eqref{eq:main3} follows as in the previous cases.

\hfill $\stackrel{\text{\small {Theorem \ref{th:fss}}}}{\Box}$

\subsection{Proof of Theorem \ref{th:bounds}}

Define preliminarily, for every $x\in[0,1]$ and $\varepsilon>0$,
\begin{eqnarray}
  E_{N,x,\varepsilon}:=\left\{ \go:\,
\frac1N \log Z_{N, \go} ^{\beta,h}\left(
 \ell_N\in [x-\varepsilon, x+\varepsilon]\right)<\frac{\tf(\beta,h)}2 \right\}.
\end{eqnarray}
Then, 
\begin{equation}
\label{eq:1Z}
\begin{split}
\bbE \left[ \frac1{Z_{N,\go}^{\beta,h}}\right] \, &\le \, 
\exp\left(-N\tf (\beta,h)/2\right)+ 
\bbE \left[\frac{\ind_{\{E_{N,x,\varepsilon}\}}}{Z_{N,\go}^{\beta,h}}\right]
\\
& \le\exp\left(-N\tf (\beta,h)/2\right)+ c_5 N^{2\alpha}\bbP[E_{N,x,\varepsilon}].
\end{split}
\end{equation}
Thanks to the Legendre 
transformation relation \eqref{eq:legendre}
and from the infinite differentiability of the free energy  for
$h<h_c(\beta)$ \cite{cf:GTloc}, it follows that the value $\bar x(h)$,
which realizes the supremum in Eq. \eqref{eq:legendre}, is unique,
smooth as a function of $h$ and satisfies $\bar x(h)=-\partial_h
\tf(\beta,h)$. Moreover, since $\phi(\beta,\bar x(h))-h \bar
x(h)=\tf(\beta,h)$, one has immediately
\begin{eqnarray}
  \lim_{N\to\infty}\frac1N \bbE \log Z_{N, \go} ^{\beta,h}\left(
\ell_N\in [\bar x(h)-\varepsilon, \bar x(h)+\varepsilon]\right)=\tf(\beta,h).
\end{eqnarray}
Thanks to Eq. \eqref{eq:conc_gen},  one has then for $\varepsilon$ sufficiently small
\begin{eqnarray}
  \bbP \left[ E_{N,\bar x(h),\varepsilon}\right] \le C_1e^{-C_2 \frac{N \tf(\beta,h)^2}{ 8\beta^2(-\partial_h\tf(\beta,h))}} 
\end{eqnarray}
for $N$ sufficiently large. Therefore, recalling Eq. \eqref{eq:1Z}, always for $N$ large one finds
\begin{equation}
\begin{split}
\label{eq:dominante}
\bbE \left[ \frac1{Z_{N,\go}^{\beta,h}}\right] \, &\le \, 
\exp\left(-N\tf (\beta,h)/2\right)+ C_1
e^{-C_2 \frac{N \tf(\beta,h)^2}{ 16\beta^2(-\partial_h\tf(\beta,h))}} 
\end{split}
\end{equation}
which immediately implies Eq. \eqref{eq:bound_mu}
for $h_c(\beta)-h>0$ sufficiently small. 
Indeed, since $\tf(\beta,.)$ is a convex function and $\tf(\beta,h_c(\beta))=0$, one has 
$$
\frac{\tf(\beta,h)}{-\partial_h\tf(\beta,h)}\le h_c(\beta)-h,
$$ 
which implies that, for $h_c(\beta)-h$ small, the second term in
the r.h.s. of Eq. \eqref{eq:dominante} is the larger  one.

\hfill $\stackrel{\text{\small {Theorem \ref{th:bounds}}}}{\Box}$

\subsection{Proof of Theorem \ref{th:correlazioni}}
\label{sec:corr}
Recall that here $\ell,k,N\in 2\N$.
We start with the upper bounds on the correlation lengths, which are
somewhat easier. Observe first that
\begin{eqnarray}
\label{eq:defC}
  \begin{split}
C_{N,\go}^{\beta,h}(\ell,k)&:= \bPo(S_\ell=S_{\ell+k}=0)-
\bPo(S_\ell=0)\bPo(S_{\ell+k}=0)\\ &=\bE_{N,\go}^{\beta,h,\otimes
2}\left[\left(\ind_{\{S^1_\ell=S^1_{\ell+k}=0\}}-\ind_{\{S^1_\ell=S^2_{\ell+k}=0\}}\right)
\ind_{\{E\}}\right],
  \end{split}
\end{eqnarray}
where $\bP_{N,\go}^{\beta,h,\otimes 2}(\cdot)$ is the product Gibbs
measure for two independent, identical copies $S^1,S^2$ of the polymer
and $E$ is the event
\begin{eqnarray}
  E=\{\nexists\, j: \ell<j<\ell+k, S^1_j=S^2_j\}.
\end{eqnarray}
Indeed, the expectation in \eqref{eq:defC} vanishes if conditioned on
the complementary of $E$, as is immediately realized via a symmetry
argument based on the Markov property of the SRW
conditioned to be non--negative.  An analogous trick was used in the
proof of  \cite[Theorem 2.2]{cf:GTloc}.  Then, it follows that
\begin{equation}
\label{eq:Clk1}
  \begin{split}
    C_{N,\go}^{\beta,h}(\ell,k)&\le \bP_{N,\go}^{\beta,h,\otimes
2}(E)=2 \bP_{N,\go}^{\beta,h,\otimes 2} \left(S^2_j>S^1_j\;\forall\,
j:\ell<j<\ell+k\right)\\ &\le 2
\bP_{N,\go}^{\beta,h}\left(S_j>0\;\forall\, j:\ell<j<\ell+k\right)
  \end{split}
\end{equation}
where in the second and third steps we used the fact that, since the polymer 
trajectories have increments
 of unit length, $S^1$ and $S^2$  cannot cross without touching.  At
this point, let us condition on the last return to zero
of $S$ before $\ell+1$, which we call $m$, and on its first return $r$
after $\ell+k-1$,  and observe that
\begin{equation}
  Z_{N,\go}^{\beta,h}\ge
Z_{N,\go}^{\beta,h}(S_m=S_r=0)=Z_{m,\go}^{\beta,h}Z_{r-m,\theta^m\go}^{\beta,h}
Z_{N-r,\theta^r\go}^{\beta,h}
\end{equation}
where, we recall, $\theta$ is the left shift: $\theta \go_n=\go_{n+1}$.
From \eqref{eq:Clk1} one obtains
\begin{equation}
  \begin{split}
\label{eq:concl1}
 C_{N,\go}^{\beta,h}(\ell,k)&\le 2\sumtwo{0\le m\le \ell}{\ell+k\le
r\le N}\bP_{N,\go}^{\beta,h}\left(\{S_m=S_r=0\}\cap \{ S_j>0\;\forall
j:m<j<r\}\right)\\ &\le 2\sumtwo{0\le m\le \ell}{\ell+k\le
r\le N}\frac{K(r-m)e^{\beta\go_r-h}}{Z_{r-m,\theta^m\go}^{\beta,h}} \le
c_6\sumtwo{0\le m\le \ell}{\ell+k\le r}\frac{1}{Z_{r-m,\theta^m\go}^{\beta,h}}.
  \end{split}
\end{equation}
Recalling the definition \eqref{eq:mu} of $\mu$ and the fact that
$(1/s)\log Z_{s,\go}^{\beta,h}$ converges to $\tf(\beta,h)$
$\bbP(\dd\go)$--a.s. for $s\to\infty$, one obtains for every
$\delta>0$
\begin{equation}
\label{eq:UB1}
  \begin{split}
     \bbE\, C_{N,\go}^{\beta,h}(\ell,k)\le c_7
     e^{-(\mu(\beta,h)-\delta)k}
  \end{split}
\end{equation}
and
\begin{equation}
\label{eq:UB2}
  \begin{split}
     C_{N,\go}^{\beta,h}(\ell,k)\le c_8(\go)
     e^{-(\tf(\beta,h)-\delta)k}
  \end{split}
\end{equation}
where $c_8(\go)$ is $\bbP(\dd \go)$--almost surely finite. Here and in
the following we omit the possible dependence on $\beta,h$ and $\ell$
of the constants, in order to keep notations lighter. 
Note 
however that $c_7$ can be chosen independent of $\ell$. Since
neither $c_7$ nor $c_8(\go)$ depend on $N$, the $N\to\infty$
limit can be taken in the l.h.s. of Eqs. \eqref{eq:UB1},
\eqref{eq:UB2}.

As for the lower bound, we start by observing that, by
Eq. \eqref{eq:defC}, one has the identity
\begin{equation}
\label{eq:newC}
  \begin{split}
    C_{N,\go}^{\beta,h}(\ell,k)= \bP_{N,\go}^{\beta,h,\otimes
2}\left(\left\{S^1_{\ell}=S^1_{\ell+k}=0\right\}\cap
\left\{S^2_j>S^1_j\; \forall\,j:\, \ell\le j\le \ell+k\right\}\right).
  \end{split}
\end{equation}
Indeed, thanks to the constraint $\ind_{\{E\}}$, it
cannot happen that $S^1_\ell=S^2_\ell=0$, otherwise also
$S^1_{\ell+1}=S^2_{\ell+1}=0$, since $S_j\ge0$ and
$|S_j-S_{j-1}|=1$. Similarly, it cannot happen that
$S^1_{\ell+k}=S^2_{\ell+k}=0$. For this reason, the first term in the
last line of \eqref{eq:defC} gives the r.h.s.  of \eqref{eq:newC}. In 
view of analogous considerations, 
 the second term is identically zero, since there are no polymer
configurations belonging to $E$, i.e., not crossing each other,  and satisfying
$S^1_\ell=S^2_{\ell+k}=0$.  On the other hand, thanks to \cite[Lemma
A.1]{cf:GTloc}, one can bound
\begin{equation}
\label{eq:lemmaA1}
  \Zno\le c_9 k^{c_9}\Zno(S_i=S_j=0)
\end{equation}
for some $c_9$ independent of $\omega$, provided that
$i,j\le 2k$.  Indeed, Lemma A.1 of \cite{cf:GTloc}
states that there exists an $\go$--independent constant
$0<c_{10}<\infty$ such that for every $N,k\in 2\N$, $k\le N$ and every
$\go$ we have 
\begin{equation}
\label{eq:A1}
  \bPo(S_k=0)\ge \frac{1}{c_{10}(k\wedge (N-k))^{c_{10}}}e^{-\beta|\go_k|-h},
\end{equation}
from which inequality \eqref{eq:lemmaA1} easily follows.

\begin{figure}[h]
\begin{center}~
\leavevmode
\epsfxsize =14 cm
\psfragscanon
\psfrag{l}[c]{{\tiny $\ell$}}
\psfrag{l-1}[c]{{\tiny $\ell\!\!-\!\!2$}}
\psfrag{l+k}[c]{{\tiny $\ell\!\!+\!\!k$}}
\psfrag{l+k+1}[c]{{\tiny $\ell\!\!+\!\!k\!\!+\!\!2$}}
\psfrag{l+logk}[c]{{\tiny $\ell\!\!+\!\!2\!\log k$}}
\psfrag{l+2logk}[c]{{\tiny $\ell\!\!+\!\!4\!\log k$}}
\psfrag{logk/2}[c]{{\tiny $\log k$}}
\psfrag{logk+1}[c]{{\tiny {\tiny $(\log k)\!\!+\!\!2$}}}
\psfrag{S1}[c]{{\tiny $S^1$}}
\psfrag{S2}[c]{{\tiny $S^2$}}
\psfrag{l-}[c]{{\tiny $\ell\!\!+\!\!k\!\!-\!\!\log k$}}
\psfrag{l+}[c]{{\tiny $\ell\!\!+\!\!\log k$}}
\psfrag{(a)}[c]{(a)}
\psfrag{(b)}[c]{(b)}
\epsfbox{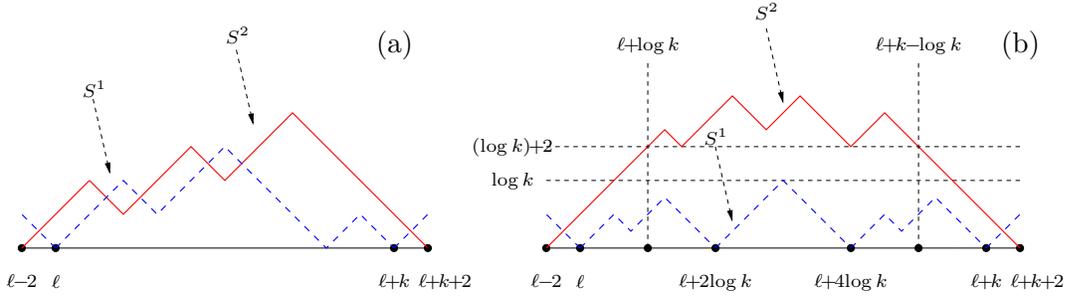}
\end{center}
\caption{\label{fig:figura} (a): Typical trajectories $S^1\in
A_1^{\ell,k}$ (dashed line) and $S^2\in A_2^{\ell,k}$ (full
line). (b): Typical trajectories $S^1\in A_3^{\ell,k}$ (dashed line)
and $S^2\in A_4^{\ell,k}$ (full line). We assumed to simplify the
picture that $\log k$ is an integer number and that $k$ is multiple of
$\log k$.  $S^2$ is constrained to go up with slope $1$ between
$\ell-2$ and $\ell+\log k$, and to go down with slope $-1$ between
$\ell+k-\log k$ and $\ell+k+2$.  Between $\ell+\log k$ and
$\ell+k-\log k$, $S^2$ cannot go below level $\log k+2=S^2_{\ell+\log
k}$.  Therefore, $S^2$ never touches zero between $\ell-1$ and
$\ell+k+1$ and $S^1$ is strictly lower than $S^2$ between $\ell$ and
$\ell+k$.  }
\end{figure}

In order to keep notations in the
following formulas simple, let us introduce some useful sets of
polymer trajectories (see Figure \ref{fig:figura}):
\begin{equation}
  \begin{split}
  A_1^{\ell,k}:=&\{S:S_\ell=S_{\ell+k}=0\}
\\ 
A_2^{\ell,k}:=&\{S:
   S_{\ell-2}=S_{\ell+k+2}=0\}
\\ 
A_3^{\ell,k}:=&\left\{S\in A_1^{\ell,k}:
   S_{\ell+j\lfloor \log k\rfloor}=0 \; \mbox{for every}\; j\in 2\N,\;1\le j\le
   \left\lfloor \frac k{\lfloor \log k\rfloor} \right\rfloor\right\}
\\
   A_4^{\ell,k}:=&\large\{S\in A_2^{\ell,k}: S_{\ell+\lfloor\log k\rfloor}=S_{\ell+k-\lfloor\log
   k\rfloor}=\lfloor\log k\rfloor+2\;\mbox{and}\;
\\ 
& S_j> \lfloor\log k\rfloor+1\;
   \mbox{for}\; \ell+\lfloor\log k\rfloor\le j\le \ell+k-\lfloor\log
   k\rfloor
\large\}.
  \end{split}
\end{equation}
Of course, $A_4^{\ell,k}$ is non--empty only for $k$ sufficiently
large so that $k\ge 2\lfloor \log k\rfloor$.
If $\hat\Omega$ is a $\bP^{\otimes 2}$--measurable set of trajectories of
$S^1,S^2$ we define, in analogy with \eqref{eq:Zvinc},
\begin{equation}
  Z_{N,\go}^{\beta,h,\otimes2}(\hat \Omega):=\bE^{\otimes2}\left(e^{\Hno(S^1)+\Hno(S^2)}
\ind_{\{(S^1,S^2)\in \hat \Omega\}}\ind_{\{S^1_N=0\}}\ind_{\{S^2_N=0\}}\right).
\end{equation}
Then, one has the obvious lower bound
\begin{equation}
  \begin{split}
\label{eq:numerator}
    C^{\beta,h}_{N,\go}(\ell,k)\ge 
\frac{
  Z_{N,\go}^{\beta,h,\otimes2}
     \left(\{S^1\in A_1^{\ell,k}\}\cap
\left\{S^2_j>S^1_j\; \forall\,j:\, \ell\le j\le \ell+k\right\}\cap
\{S^2\in A_2^{\ell,k}\}
\right)
}{(\Zno)^2}
  \end{split}
\end{equation}
and, thanks to Eq. \eqref{eq:lemmaA1}, 
\begin{equation}
  (\Zno)^2\le c_9^2 k^{2c_9}\Zno(S\in A_1^{\ell,k})\Zno(S\in A_2^{\ell,k}).
\end{equation}
The numerator in \eqref{eq:numerator} can be bounded below
requiring that $S^1\in A_3^{\ell,k}$ and $S^2\in A_4^{\ell,k}$.  At
this point the constraint $\{S^2_j>S^1_j\; \forall\,j:\, \ell\le j\le
\ell+k\}$ becomes superfluous, since it is automatically satisfied if
$S^1\in A_3^{\ell,k}$ and $S^2\in A_4^{\ell,k}$, and one obtains
  \begin{equation}
\label{eq:plug}
    \begin{split}
 C^{\beta,h}_{N,\go}(\ell,k)\ge 
c_9^{-2}k^{-2c_9}\frac{Z_{k,\theta^{\ell}\go}^{\beta,h}(S\in A_3^{0,k})}{Z_{k,\theta^{\ell}\go}^{\beta,h}}
\frac{Z_{k+4,\theta^{\ell-2}\go}^{\beta,h}(S\in A_4^{2,k})}{Z_{k+4,\theta^{\ell-2}\go}^{\beta,h}}.
    \end{split}
  \end{equation}
Note that the trajectories belonging to $A_4^{2,k}$ never touch the
defect line in the interval $\{1,\ldots,k+3\}$. Therefore, in
$Z_{k+4,\theta^{\ell-2}\go}^{\beta,h}(S\in A_4^{2,k})$ the pinning
Hamiltonian gives no contribution except at the boundary point $k$,
and one is left with a SRW computation.  An easy
counting of allowed trajectories gives, for large $k$,
\begin{equation}
\label{eq:f1}
  Z_{k+4,\theta^{\ell-2}\go}^{\beta,h}(S\in A_4^{2,k})\ge
  k^{-c_{11}}
\end{equation}
uniformly in $\go$. Secondly, applying repeatedly \cite[Lemma
A.1]{cf:GTloc} one obtains
\begin{equation}
\label{eq:f2}
  Z_{k,\theta^{\ell}\go}^{\beta,h}(S\in A_3^{0,k})\ge c_{12}^{-k/\log k}\left(\log k\right)^{-c_{12}k/\log k}
Z_{k,\theta^\ell \go}^{\beta,h}.
\end{equation}
Plugging the lower bounds \eqref{eq:f1}, \eqref{eq:f2} into
\eqref{eq:plug} and taking the $N\to\infty$ limit one finally finds
\begin{equation}
 C^{\beta,h}_{\infty,\go}(\ell,k)\ge \frac{c_{13}
e^{-c_{14}\frac{k}{\log k}\log(\log k)}}
{Z_{k+4,\theta^\ell\go}^{\beta,h}}.
\end{equation}
The conclusions 
\begin{equation}
\bbE\,  C_{\infty,\go}^{\beta,h}(\ell,k)\ge c_{15}e^{-(\mu(\beta,h)+\delta)k}
\end{equation}
and
\begin{equation}
  C_{\infty,\go}^{\beta,h}(\ell,k)\ge c_{16}(\go)e^{-(\tf(\beta,h)+\delta)k}
\end{equation}
are obtained, for every $\delta>0$, by
recalling the definition of $\mu(\beta,h)$ and the fact that
$(1/k)\log Z_{k,\go}^{\beta,h}$ converges to $\tf(\beta,h)$ almost
surely. Together with Eqs. \eqref{eq:UB1}, \eqref{eq:UB2}, these
imply the desired results \eqref{eq:corr_ave}, \eqref{eq:corr_typ}.

\hfill $\stackrel{\text{\small {Theorem \ref{th:correlazioni}}}}{\Box}$

\begin{rem}\rm 
  It is interesting to compare the strategy leading to the upper
  bounds \eqref{eq:UB1}, \eqref{eq:UB2} with the coupling method
  introduced in Ref.  \cite{cf:LT} to estimate the speed of
  convergence to equilibrium of some special renewal sequences.  The
  connection between polymer measures and renewal equations is not
  casual: for instance, a moment of reflection (or a look at Appendix
  A of \cite{cf:GT05}) shows that, in the homogeneous case $\beta=0$,
  the polymer measure can be rewritten exactly in terms of the renewal
  process where the probability that the time elapsed between two
  successive renewals is $n$ is given by $K(n)\exp(-n
  \tf(0,h)-h)$.

\end{rem}

\section*{acknowledgments} 
I am extremely grateful to Giambattista Giacomin for countless
conversations on these topics, and in particular about the contents of
Section \ref{sec:xi}.  
 Most of this work was written during a stay at
 the Isaac Newton Institute in Cambridge, in the framework of the
 programme ``Principles of the Dynamics of Non-Equilibrium Systems''.
This work was partially supported by the GIP--ANR project JC05\_42461
({\sl POLINTBIO}).

\end{document}